\documentclass[conference]{IEEEtran}
\IEEEoverridecommandlockouts

\usepackage{cite}
\usepackage{amsmath,amssymb,amsfonts}
\usepackage{graphicx}
\usepackage{textcomp}
\usepackage{xcolor}
\usepackage{url}
\usepackage{hyperref}
\usepackage{booktabs}
\usepackage{float}
\usepackage{placeins}

\def\BibTeX{{\rm B\kern-.05em{\sc i}\kern-.025em b}\kern-.08em\TeX}

\begin{document}

\title{Integrating Delay-Absorption Capability into Flight Departure Delay Prediction}

\author{\IEEEauthorblockN{Jianyang Zhou}
\IEEEauthorblockA{\textit{George Mason University} \\
Fairfax, VA \\
jzhou30@gmu.edu}
}

\maketitle

\begin{abstract}
Accurately forecasting flight departure delays is essential for improving operational efficiency and mitigating the cascading disruptions that propagate through tightly coupled aircraft rotations. Traditional machine learning approaches often treat upstream delays as static variables, overlooking the dynamic recovery processes that determine whether a delay is absorbed or transmitted to subsequent legs. This study introduces a two-stage machine learning framework that explicitly models delay-absorption behavior and incorporates it into downstream delay prediction. In Stage~I, a CatBoost classifier estimates the probability that a flight successfully absorbs an upstream delay based on operational, temporal, and meteorological features. This probability, termed \textit{AbsorbScore}, quantifies airport- and flight-specific resilience to delay propagation. In Stage~II, an XGBoost classifier integrates AbsorbScore with schedule, weather, and congestion indicators to predict whether a flight will depart more than 15 minutes late. Using U.S. domestic flight and NOAA weather data from Summer 2023, the proposed framework achieves substantial improvements over baseline models, increasing ROC--AUC from 0.865 to 0.898 and enhancing precision to 89.2\% in identifying delayed flights. The results demonstrate that modeling delay absorption as an intermediate mechanism significantly improves predictive performance and yields interpretable insights into airport recovery dynamics, offering a practical foundation for data-driven delay management and proactive operational planning.

\end{abstract}

\begin{IEEEkeywords}
Flight Delay Prediction, Delay Propagation, CatBoost, XGBoost, Weather, Airport, Machine Learning
\end{IEEEkeywords}

\section{Introduction}
Flight delays have long been a critical concern in the aviation industry, posing significant challenges to airline operations, passenger satisfaction, and overall economic performance. Each year, delays cause substantial financial losses for airlines, disrupt operational schedules, and negatively affect passenger experiences. According to the U.S. Department of Transportation, flight delays cost the U.S. economy billions of dollars annually through increased airline operational expenses and the economic impact of lost passenger time and missed connections. As air traffic networks become increasingly complex and travel demand continues to grow, accurately predicting and managing flight delays has become essential for improving operational efficiency and mitigating disruptions.

Traditionally, delay prediction has relied on statistical analysis and rule-based approaches that often fail to capture the nonlinear and dynamic nature of air transportation systems. Recent advancements in machine learning (ML) and deep learning (DL) have shown great promise in modeling complex temporal and spatial dependencies in flight data. 

However, an important aspect frequently overlooked in delay prediction research is an airport’s or flight’s ability to absorb upstream delays—for example, when an inbound aircraft arrives late but still enables an on-time outbound departure due to scheduling flexibility or efficient ground handling. Understanding and quantifying this delay absorption capacity is crucial for distinguishing between transient and systemic delay propagation, thereby improving both prediction reliability and operational insight.

To address this gap, this paper proposes a novel two-stage machine learning framework. In the first stage, a classification model predicts whether a flight successfully absorbs an upstream delay, represented by the binary variable DelayAbsorbed. In the second stage, the predicted probability (AbsorbScore) is incorporated as a contextual feature in a regression model for departure delay prediction. This framework jointly captures the resilience of airport operations and the dynamic interactions between sequential flights.

By integrating flight attributes, weather conditions, and airport congestion metrics with learned delay absorption behavior, the proposed approach aims to enhance both predictive accuracy and interpretability. The findings provide actionable insights for airlines and airport operators to design more robust schedules, allocate turnaround resources efficiently, and implement data-driven delay management strategies that improve overall system performance and passenger experience.

\section{Literature Review}

\subsection{Machine Learning Approaches}

Flight delays are a critical issue that impacts the economy, passenger satisfaction, and airline operations. With growing air traffic and interconnected networks, predicting and mitigating delays has become increasingly important. Early research about using machine learning to predict flight delay can be tracked back to 2008, where $k$-Means clustering was applied to analyze historical flight data, identifying five delay grades of delay ---from no delay to heavy delay, based on similarities in key metrics \cite{b1}. 

Recent years have witnessed significant growth in the application of machine learning techniques to predict flight delays. Esmaeilzadeh and Mokhtarimousavi employed a Support Vector Machine (SVM) model, utilizing data such as flight schedules, airport demand/capacity, and weather conditions from three major New York City airports, achieving 85.57\% accuracy for arrival delay prediction \cite{b2}. Li and Jing applied a random forest model using features such as airport congestion, network congestion, and demand-capacity imbalance to U.S. domestic flight data from July 2018, achieving arrival delay prediction errors smaller than 30 minutes in 97\% of cases \cite{b3}. Yu et al. introduced a deep belief network for Beijing International Airport data, incorporating explanatory variables such as prior flight delays, airline characteristics, aircraft capacity, and boarding options, which delivered high prediction accuracy \cite{b4}. Guo et al. developed a hybrid technique combining Random Forest Regression with the Maximal Information Coefficient, using inputs such as aircraft capacity, prior delays, turnaround times, and airport crowding, outperforming traditional methods like linear regression and neural networks \cite{b5}. Similarly, Gui et al. \cite{b6} evaluated LSTM and Random Forest models with inputs like traffic flow, weather, flight schedules, and ADS-B signals, identifying Random Forest as the most efficient approach.

Despite recent progress, existing models still struggle to fully capture the temporal dependencies inherent in sequential flight operations. Rebollo and Balakrishnan~\cite{b7,b8} and Gopalakrishnan and Balakrishnan~\cite{b9} constructed delay propagation networks based on origin–destination (OD) pairs, providing valuable insight into network-wide delay transmission. Similarly, Güvercin, Ferhatosmanoglu, and Gedik~\cite{b10} and Li and Jing~\cite{b11} represented the air traffic system as a graph of nodes and edges to extract temporal characteristics of delay spread across airports. However, these network-based approaches often treat delay propagation as discrete and static, overlooking the continuous and dynamic evolution of flight sequences.

\subsection{Delay Propagation in Air Transportation}

Flight delay propagation refers to the process by which an initial delay in one flight spreads to subsequent operations through shared resources such as aircraft rotations, crew schedules, and airport congestion. Since a single aircraft typically performs multiple flight legs per day, a small upstream delay can cascade through downstream flights, amplifying system-wide disruption. Understanding how delays propagate and how buffers mitigate these effects is essential for modeling airport performance and improving delay prediction.

\subsubsection{Analytical Foundations}
Early analytical work by Boswell and Evans \cite{b12} developed a Markov-based framework to describe \textit{downstream delay propagation}. Their model quantified how an upstream departure delay affects subsequent arrivals and departures within an aircraft rotation, showing that each flight segment inherits a portion of prior delay depending on recovery opportunities and operational uncertainty. Although pioneering, this framework primarily captured temporal propagation and did not explicitly model how \textit{buffers}—either airborne or ground-based—absorb delays.

\subsubsection{Analytical--Econometric Framework}
Kafle and Zou (2016) \cite{b4} extended this concept by proposing a hybrid analytical--econometric model that decomposes the total observed delay at a node $i$ into two components:
\begin{equation}
O_i = N_i + \sum_{k=1}^{i-1} p_{k,i},
\end{equation}
where $N_i$ represents the \textit{newly formed delay} and $p_{k,i}$ denotes the \textit{propagated delay} inherited from previous legs.
Their \textit{time--space graph} representation models each flight as a link between two nodes (departure and arrival), introducing both \textit{flight (block) buffers} and \textit{ground (turnaround) buffers} as mechanisms that absorb portions of delay.
The authors proposed three buffer allocation scenarios—buffers first absorbing newly formed delays, propagated delays, or both proportionally—to capture different operational realities.
An econometric model based on Heckman’s two-step estimation was used to quantify how buffer size, congestion level, and airport characteristics influence the initiation and progression of propagated delays.
This framework provided one of the first quantitative evaluations of how buffer management mitigates delay transmission across the U.S. air transportation network.

\begin{figure*}
\centering
\includegraphics[width=0.95\linewidth]{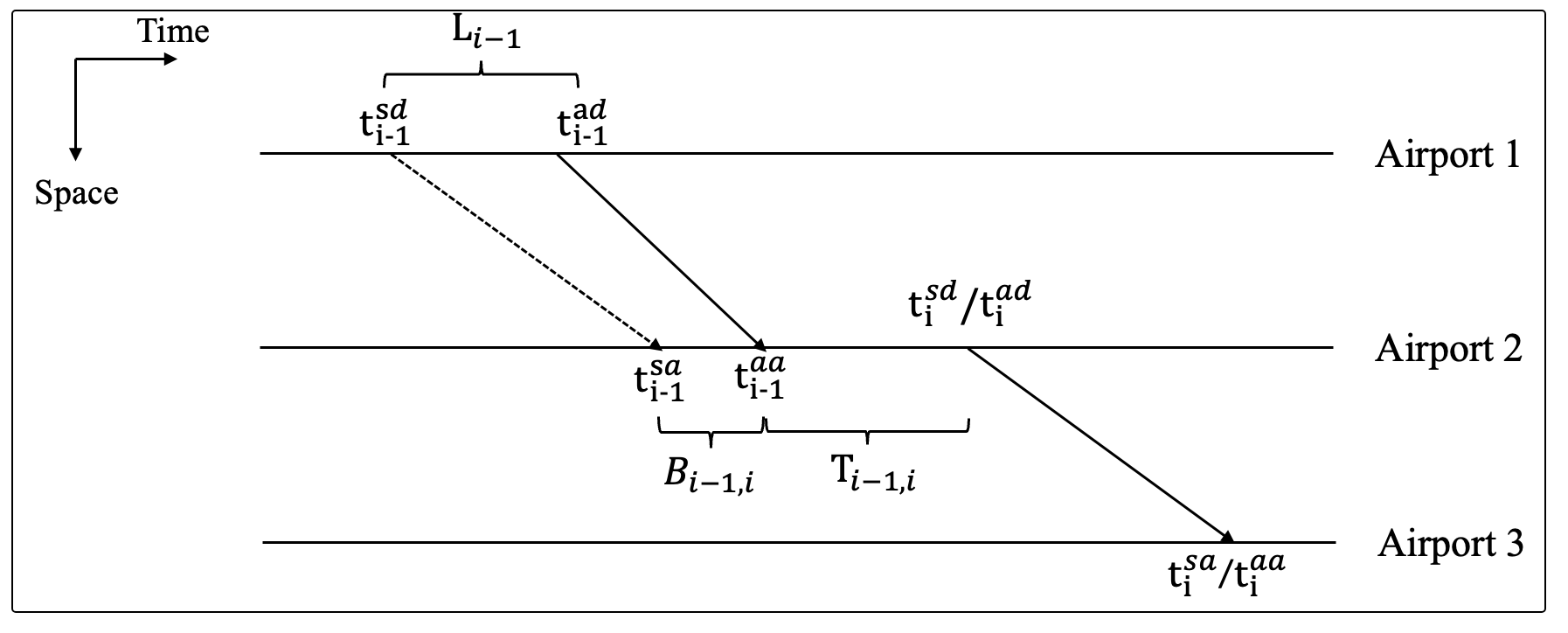}
\caption{Time-space graph of a single flight and ground buffer time illustration}
\label{fig:buffer}
\end{figure*}

Figure~\ref{fig:buffer} illustrates a time--space diagram for two consecutive flight legs---flight $i-1$ from Airport~1 to Airport~2, followed by flight $i$ from Airport~2 to Airport~3. 
The horizontal axis represents time, while the vertical axis represents spatial position across the three airports.

For flight $i-1$, the scheduled departure and arrival times are denoted as $t^{sd}_{i-1}$ and $t^{sa}_{i-1}$, respectively, whereas the actual departure and arrival times are $t^{ad}_{i-1}$ and $t^{aa}_{i-1}$. 
The deviations between scheduled and actual timelines capture any upstream delay that may propagate into downstream operations.

Once flight $i-1$ arrives at Airport~2 at time $t^{aa}_{i-1}$, the aircraft begins its turnaround process in preparation for flight $i$. 
The scheduled departure time of flight $i$ is $t^{sd}_{i}$, and the scheduled arrival time at Airport~3 is $t^{sa}_{i}$ (with actual times $t^{ad}_{i}$ and $t^{aa}_{i}$).

The \textbf{ground buffer} between flights $i-1$ and $i$, denoted $B_{i-1,i}$, is represented by the time interval between the aircraft’s \textit{scheduled arrival} from flight $i-1$ and the minimum required turnaround duration prior to the departure of flight $i$. 
This buffer quantifies the slack time available to absorb delays from the previous leg. 
If flight $i-1$ arrives late but the delay is smaller than $B_{i-1,i}$, the turnaround can still be completed on time, allowing flight $i$ to depart according to schedule. 
Conversely, if the upstream delay exceeds this buffer, the excess delay propagates and causes a departure delay for flight $i$.

The actual turnaround time $T_{i-1,i}$ is shown as the interval between the actual arrival $t^{aa}_{i-1}$ and the actual departure $t^{ad}_{i}$. 
The interaction among scheduled arrival, realized arrival delay, ground buffer, and turnaround processes determines whether delay is absorbed or propagated to the next flight leg.

\subsubsection{Network-Level Classification}
Kim and Park \cite{b13} further generalized the analytical framework to a network-level classification model using large-scale South Korean flight operation data.
By distinguishing \textit{newly formed}, \textit{propagated}, and \textit{generated} delays, their study identified airports' dual roles as delay absorbers or amplifiers.
Using multiple regression on route-level and airport-level features (e.g., block time, turnaround time, and congestion metrics), they quantified each airport’s \textit{delay absorption efficiency}.
Two derived indicators—\textit{Average Generated Delay (AGD)} and \textit{Average Propagated Delay (APD)}—were used to classify airports into four categories:
(A) high-generation/high-propagation hubs, 
(B) low-generation/high-propagation nodes,
(C) high-generation/low-propagation airports, and 
(D) robust low-delay airports.
This macro-level classification revealed how airport-specific scheduling buffers and operational practices shape delay diffusion across the network. However, these approaches do not explicitly model whether a delay is absorbed during turnaround, a key mechanism driving downstream delay propagation.

\subsection{Research Gap and Contribution}

Existing studies on flight delay prediction have primarily focused on estimating the occurrence or duration of delays at the flight level using statistical or machine learning techniques. 
Models such as random forests, gradient boosting algorithms (e.g., XGBoost, CatBoost, LightGBM), and deep neural networks have demonstrated strong predictive performance when incorporating features related to weather, scheduling, and air traffic conditions. 
However, most of these models treat delay propagation mechanisms---particularly the link between previous flight delays, turnaround processes, and subsequent departures---as static input features rather than dynamic temporal relationships. 
As a result, they often overlook the \textit{delay absorption process}, in which ground operations recover from upstream delays and restore schedule regularity before the next flight cycle.

From an operational and transportation economics perspective, prior research has shown that airlines intentionally introduce schedule buffers and turnaround slack to mitigate reactionary delays, which account for a substantial portion of total departure delays. 
While these studies provide valuable theoretical and empirical insights into delay propagation and recovery mechanisms, they generally operate at an aggregated level---airport, route, or network---and rarely translate these mechanisms into \textit{learnable, flight-level representations}. 
Specifically, the conditional relationships among previous arrival delay, turnaround duration, and subsequent departure delay are often simplified into static indicators, such as average buffer time or airport dummies, rather than being modeled dynamically within a data-driven framework.

Recent research has begun to examine the role of scheduling flexibility and turnaround management in reducing delay propagation, yet a unified modeling framework remains limited. 
Specifically, there is a lack of approaches that (i) explicitly quantify the probability of \textit{delay absorption} at the individual flight level; (ii) integrate this learned probability into downstream delay prediction models; and (iii) evaluate its transferability and interpretability across heterogeneous airport and weather contexts.

Although existing analytical and econometric frameworks---such as those developed by Boswell and Evans~\cite{b12}, Kafle and Zou~\cite{b14}, Zheng et al.~\cite{b13}, and Kim and Park~\cite{b15}---offer important theoretical foundations, they rely on static or parametric assumptions regarding buffer efficiency. 
In reality, an airport’s ability to absorb delay varies dynamically with factors such as operational load, weather, and scheduling configurations. 

To address these limitations, the present study proposes a \textbf{two-stage machine learning framework} that explicitly models airport-level delay absorption behavior. 
In the first stage, the model learns the probability that an upstream delay is absorbed during turnaround operations (\textit{DelayAbsorbed}). 
In the second stage, this probabilistic output (\textit{AbsorbScore}) is incorporated into a flight-level prediction model to enhance both the accuracy and interpretability of departure delay forecasts. 
By quantifying each airport’s dynamic delay absorption capability---both during ground turnaround and in airborne phases---this study advances a more realistic and adaptive understanding of network-level delay resilience and recovery dynamics.

\section{Dataset Review and Feature Engineering}
\subsection{Dataset Overview}
We obtained flight data from the Bureau of Transportation Statistics (BTS) \textit{Airline On-Time Performance} database, which provides monthly records of domestic flights within the United States~\cite{b16}. 
Each monthly dataset contains approximately 500{,}000 flight records with 123 attributes, including departure and arrival airports, gate times, airborne times, and categorized delay causes, as detailed in Appendix~A. 
We manually downloaded flight data from June 2023 to August 2023 and merged them into a single dataset for this study.

Weather data were collected from the National Oceanic and Atmospheric Administration (NOAA) official database, which reports hourly meteorological conditions at airports across the United States~\cite{b17}. 
We manually downloaded weather data for 342 airports. 
Each record includes up to 140 variables describing meteorological factors such as temperature, wind speed, wind direction, precipitation, and visibility, as detailed in Appendix~B.

The input features consist of four categories:

\begin{itemize}
    \item \textbf{Categorical features:} operating airline, origin and destination airports, month, and scheduled departure time block.
    \item \textbf{Temporal features:} day of week, scheduled times of departure and arrival, and departure hour.
    \item \textbf{Operational features:} previous flight delay, turnaround time, flight distance, and their interactions 
    (e.g., departure hour $\times$ low-visibility indicator, turnaround time $\times$ weather severity).
    \item \textbf{Weather features:} hourly wind speed, precipitation, visibility, and humidity at the departure airport. 
    These are standardized and combined into a composite weather severity index.
\end{itemize}
We derive features from scheduled departure time (converted to hour-of-day), carrier, airport codes, flight distance, and day of the week. We also construct a weather severity index (WxSeverity) combining normalized wind speed, precipitation, visibility (inverted), and humidity. Additional interaction terms include departure hour $\times$ low visibility flag and turnover buffer $\times$ weather severity.

Because BTS data do not explicitly record aircraft rotations, we reconstruct the previous flight leg for each operation using \textit{Tail Number} and chronological sequencing. Flights are grouped by tail number and sorted by date and scheduled departure time. For each flight, the preceding leg is defined as the most recent flight performed by the same aircraft whose destination matches the current flight's origin and whose actual arrival precedes the scheduled departure. This procedure ensures a physically feasible chain of operations and provides accurate computation of upstream delay and turnaround buffers, which are required for defining \textit{DelayAbsorbed} in Stage~I.

\subsection{Departure Delay Characteristics}

\begin{figure}[H]
\centering
\includegraphics[width=0.95\linewidth]{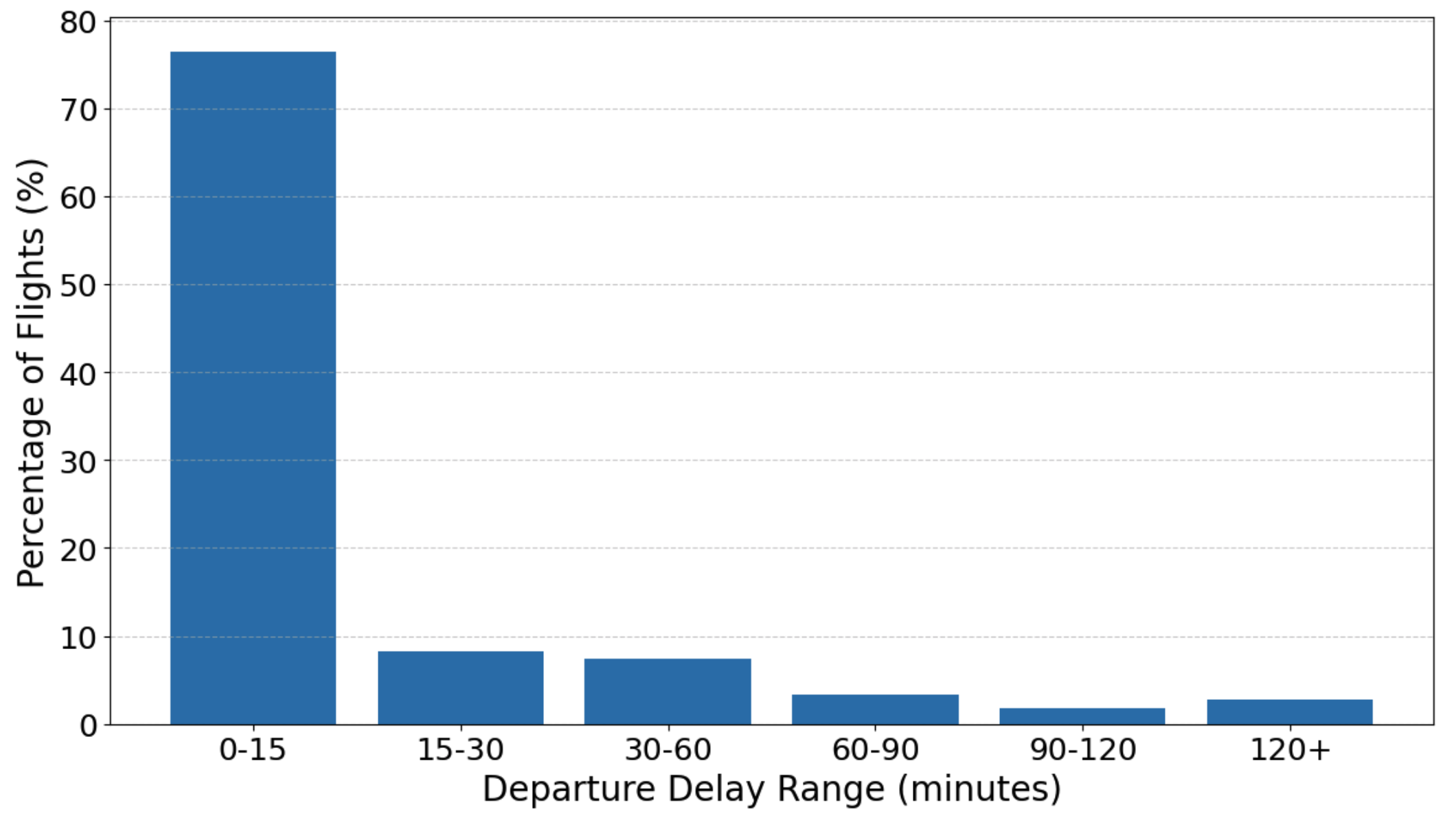}
\caption{Distribution of departure delay categories}
\label{fig:dep_delay_stat}
\end{figure}

The dataset exhibits moderate departure delay behavior. The average departure delay across all flights is \textbf{13.01 minutes}, indicating that most operations experience only minor schedule deviations. However, a substantial share of flights shown in Figure~\ref{fig:dep_delay_stat} still encounter more meaningful disruptions: \textbf{22.77\%} of all departures leave more than 15 minutes behind schedule, which is the standard threshold commonly used in the aviation industry to classify a flight as ``delayed.'' 

\subsection{Airport's performance on delay absorption}

\begin{figure}[H]
\centering
\includegraphics[width=0.95\linewidth]{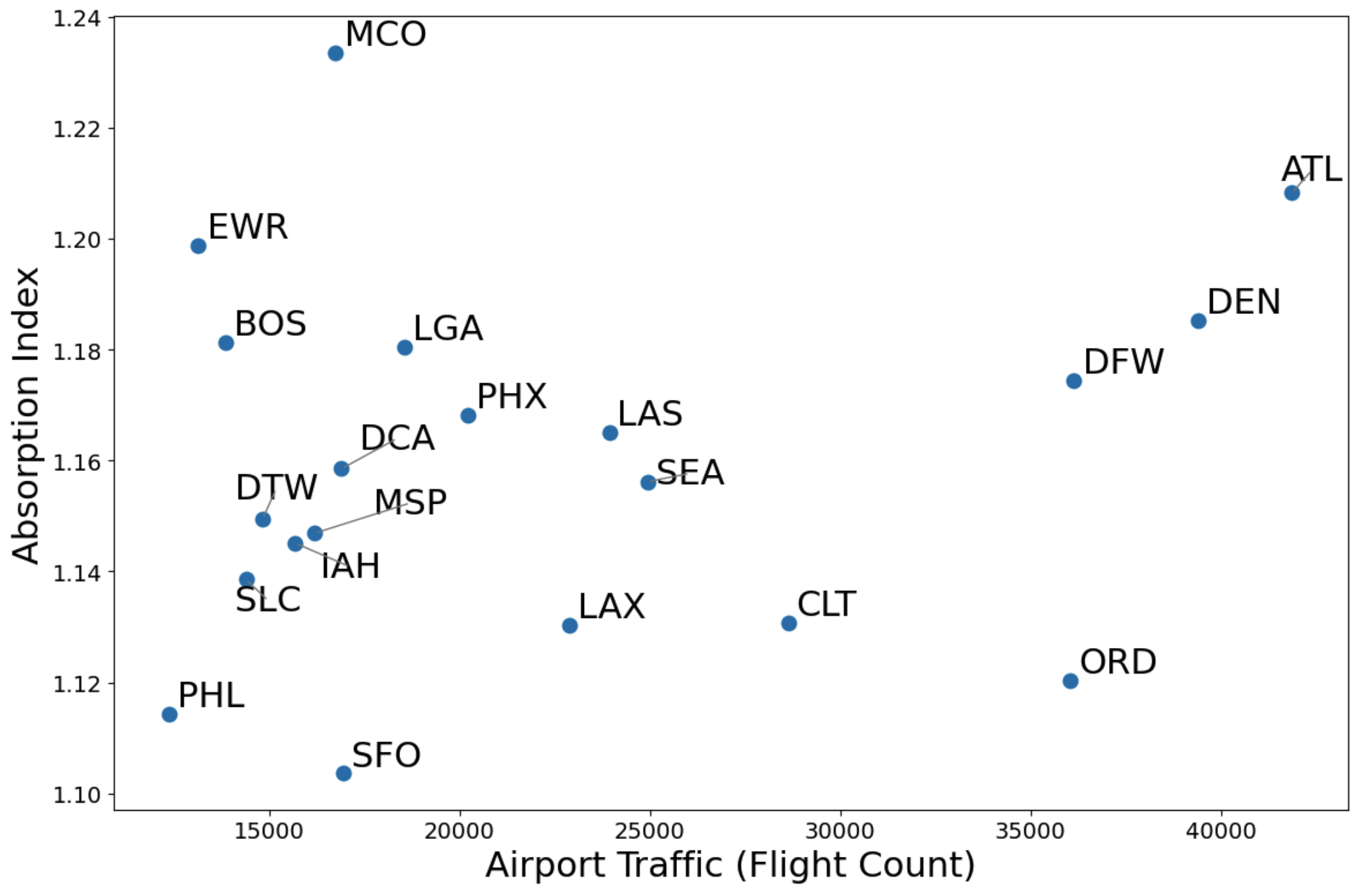}
\caption{Top 20 busiest airports: traffic vs delay absorption index}
\label{fig:airport_stat}
\end{figure}

Figure~\ref{fig:airport_stat} illustrates how airport traffic volume relates to the average Turnover Index, which is computed as the ratio between Actual Turnover Time and Scheduled Turnover Time for all flights operating at each airport. Specifically, the Turnover Index for flight $i$ is defined as
\[
\text{Turnover Index}_i = \frac{\text{ActualTurnover}_i}{\text{ScheduledTurnover}_i}.
\]
To obtain this result, flights involving overnight ground time (defined as aircraft remaining on the ground for more than five hours) were removed to ensure that the analysis reflected true same-day turnaround performance rather than overnight layovers. For each airport, the Turnover Index was calculated at the flight level and then averaged across all valid operations, while airport traffic was measured as the total number of flights handled within the dataset. Plotting these two metrics together allows us to visualize how busy airports compare with less busy ones in their ability to recover schedule delays during ground operations.

Overall, the plot shows no strong linear relationship between traffic volume and delay-absorption performance. Several high-traffic hubs such as ATL, DEN, and DFW exhibit relatively high Turnover Index values (above 1.18), indicating that they tend to operate with longer-than-scheduled turnaround times and therefore have limited ability to absorb upstream delays. In contrast, airports such as ORD, CLT, LAX, and SFO display lower Turnover Index values (around 1.10--1.13), suggesting stronger delay-recovery capability despite handling moderate to high traffic. Among medium-traffic airports, performance varies substantially: MCO and EWR demonstrate relatively poor absorption (approximately 1.20 or higher), while PHL achieves the strongest absorption efficiency (around 1.11). Taken together, these patterns indicate that delay-absorption performance is influenced more by airport-specific operational characteristics than by traffic volume alone.

To investigate how the magnitude of upstream delay affects an airport's delay-absorption capability, we categorize the previous flight's arrival delay into five levels: 0--15, 15--30, 30--60, 60--90, and 90--120 minutes. For each delay category, we compute the corresponding average absorbed delay. The absorbed delay is defined as the difference between the upstream arrival delay and the downstream departure delay, i.e.,
\[
\text{AbsorbedDelay}_i = \text{PreviousDelay}_i - \text{DepDelay}_i.
\]

\begin{figure}[H]
\centering
\includegraphics[width=0.95\linewidth]{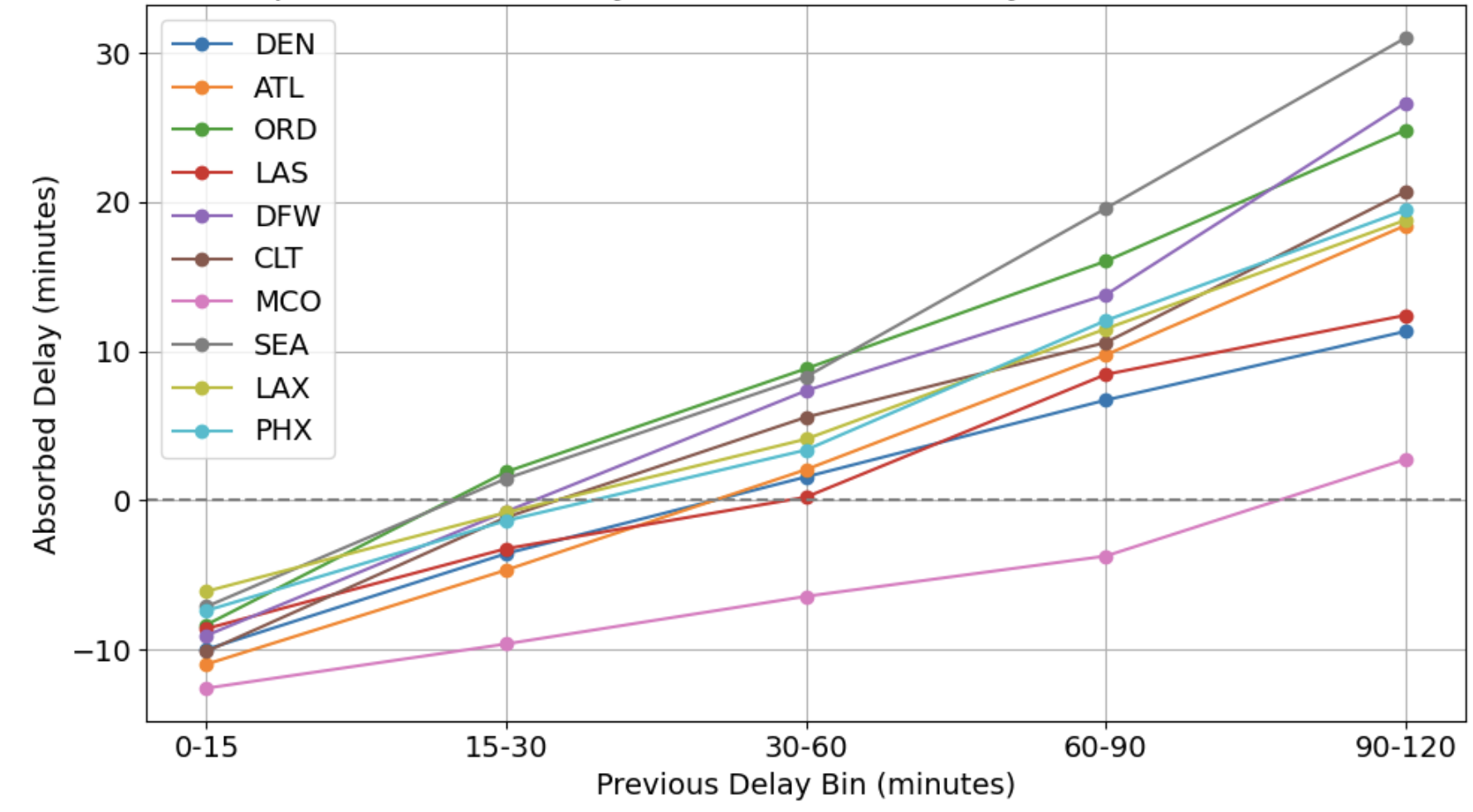}
\caption{Airport absorbed delay across previous delay levels}
\label{fig:airport_stat_2}
\end{figure}

Figure~\ref{fig:airport_stat_2} summarizes how major U.S. airports absorb upstream delays across different levels of previous delay (0–120 minutes). For small upstream delays (0–15 minutes), all airports exhibit negative absorbed delay values, indicating that they tend to amplify rather than recover minor delays. As previous delays increase, most airports transition into positive absorbed delay, showing increasing capability to recover a portion of the inbound delay. Among the airports shown, SEA, ORD, and DFW consistently demonstrate the strongest delay-absorption behavior, recovering over 20 minutes of delay when previous delay exceeds 90 minutes. In contrast, MCO exhibits notably weak absorption across most delay ranges, consistently amplifying rather than mitigating delays. Overall, the figure reveals that delay-recovery performance varies substantially by airport and improves as inbound delays become larger, although the degree of improvement is highly airport-dependent.

To further investigate how adverse weather affects an airport’s ability to absorb upstream delays, we examine the absorbed-delay profiles of Atlanta Hartsfield–Jackson International Airport (ATL) on the two operationally most contrasting days in our dataset. On 6 August 2023, ATL recorded the highest average departure delay of 61.6 minutes, coinciding with a period of heavy rain and mist in the afternoon that significantly degraded runway throughput and ground-movement efficiency. In contrast, 22 August 2023 exhibited the lowest average departure delay of only 1.3 minutes under clear and stable meteorological conditions. These two dates provide a natural experiment for assessing weather-driven variations in delay-recovery performance.

\begin{figure}[H]
\centering
\includegraphics[width=0.95\linewidth]{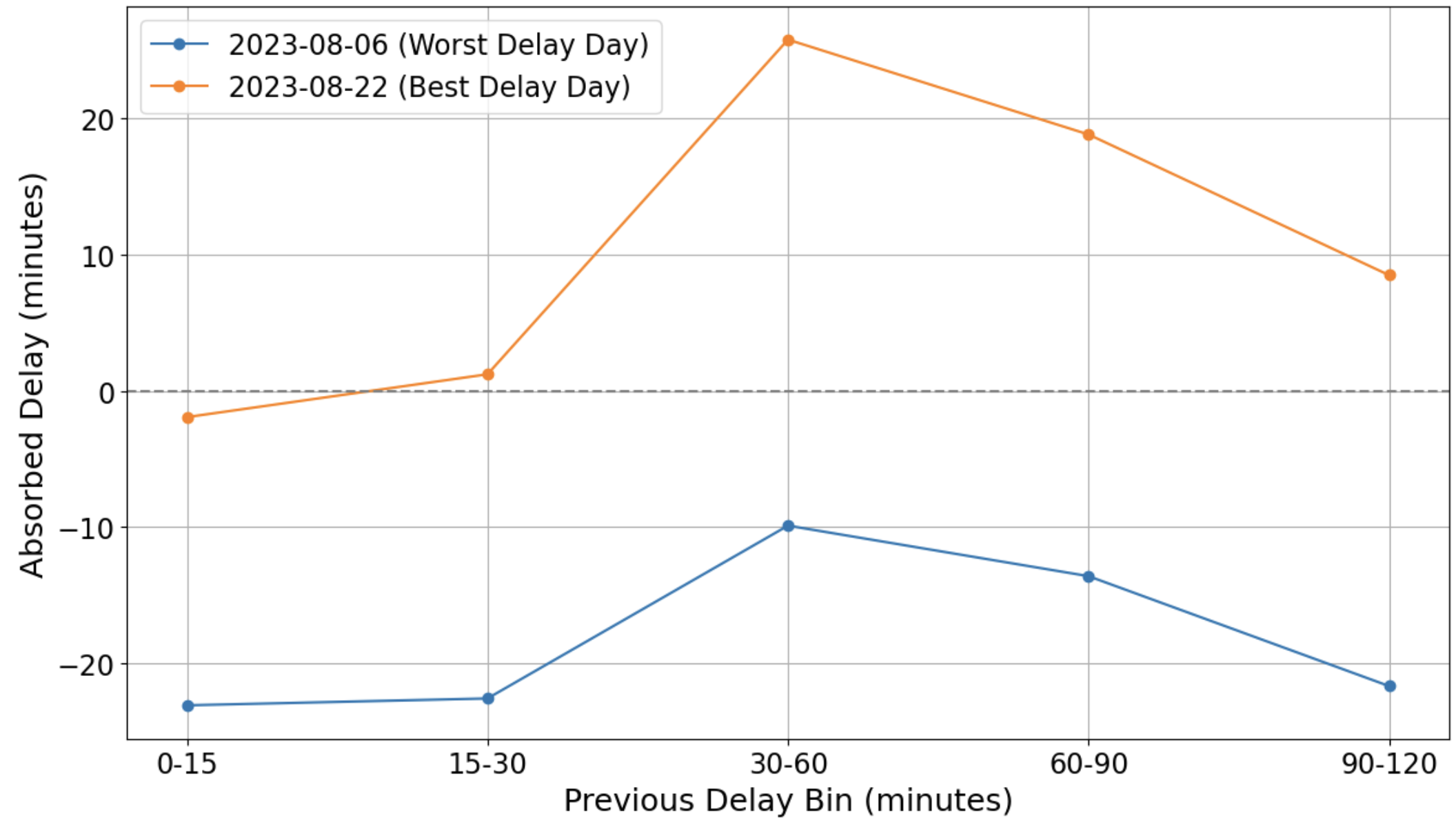}
\caption{ATL's delay absorption capability in different weather}
\label{fig:airport_stat_3}
\end{figure}

Figure~\ref{fig:airport_stat_3} compares ATL’s absorbed delay across different ranges of upstream delay (0–120 minutes) for the two days. Under adverse weather on 6 August, ATL consistently amplified inbound delays across all bins, with absorbed delay values ranging from roughly $-25$ to $-5$ minutes, indicating an inability to recover delays even when upstream delays exceeded 60 minutes. Conversely, on 22 August, ATL exhibited positive absorbed delay—recovering between 2 and 25 minutes depending on the upstream delay level—demonstrating effective delay mitigation capability under favorable weather conditions. These contrasting patterns highlight the strong sensitivity of airport delay-absorption performance to local weather: heavy precipitation and reduced visibility substantially diminish an airport’s operational flexibility and its capacity to compress turnaround processes, whereas clear weather enables more efficient recovery from upstream disruptions.

These results confirm that delay absorption is highly airport-dependent and non-linear with respect to upstream delays, which motivates the need for a learning-based absorption model.

\section{Methodology}

\subsection{Overview}
To capture both the propagation and absorption characteristics of flight delays, we develop a two-stage learning framework that integrates operational, temporal, and meteorological information from flight and weather datasets. The overall workflow consists of: (1) constructing an intermediate model to quantify each flight's delay absorption capability, and (2) incorporating this measure into a final model that predicts departure delay occurrence. This design allows the model to explicitly encode how previous delays propagate through aircraft rotations while accounting for airports' heterogeneous capacities to recover from disruptions.

\subsection{Definition of \texttt{DelayAbsorbed}}

To determine whether a flight successfully mitigates the delay inherited from its upstream leg, we adopt a buffer-based definition that accounts for systematic differences in turnaround performance across airports and carriers. Turnaround efficiency depends on local ground-handling practices, resource availability, and airline scheduling philosophy; thus, a universal threshold cannot capture this operational heterogeneity. Instead, we estimate the typical schedule buffer available to each airport--carrier pair.

Let $\mathcal{F}_{a,c}$ denote the set of flights that turn around at airport $a$ and are operated by carrier $c$. For each $j \in \mathcal{F}_{a,c}$, let $T^{\text{scheduled}}_j$ and $T^{\text{actual}}_j$ denote the scheduled and realized turnaround durations. The \emph{average buffer time} for the pair $(a,c)$ is defined as
\[
\overline{B}_{a,c}
= \frac{1}{|\mathcal{F}_{a,c}|}
  \sum_{j \in \mathcal{F}_{a,c}}
  \bigl(T^{\text{scheduled}}_j - T^{\text{actual}}_j \bigr),
\]
representing the baseline amount of delay this airport--carrier combination typically absorbs under routine operations.

For a specific flight $i$, the realized recovery of inherited delay is computed as
\[
\Delta_i = D_{\text{prev},i} - D_{\text{dep},i},
\]
which measures how many minutes of upstream delay are eliminated during turnaround. A flight is classified as having successfully absorbed its upstream delay if its realized recovery meets or exceeds the historical buffer capability of its corresponding airport--carrier pair:
\[
\texttt{DelayAbsorbed}_i =
\begin{cases}
1, & \text{if } \Delta_i \ge \overline{B}_{a,c}, \\
0, & \text{if } \Delta_i < \overline{B}_{a,c}.
\end{cases}
\]
This definition provides a behaviorally grounded, context-sensitive indicator of delay absorption that reflects each airport--carrier pair’s typical operational performance.

\subsection{Stage I: Modeling Delay Absorption}

Stage~I aims to model the binary absorption outcome defined above. For each flight with positive inherited delay ($D_{\text{prev}}>0$), the label \texttt{DelayAbsorbed} indicates whether the realized recovery $\Delta_i$ is sufficiently large relative to the airport--carrier specific benchmark $\overline{B}_{a,c}$. Because this absorption behavior depends on a wide range of operational and environmental factors, we train a CatBoost classifier to predict
\[
P(\texttt{DelayAbsorbed}_i = 1 \mid \mathbf{x}_i),
\]
where $\mathbf{x}_i$ includes airport and carrier identifiers, turnaround-related features, temporal descriptors (month, departure time block, departure hour), and NOAA-derived weather indices.

The classifier output is stored as the \textbf{AbsorbScore}, a continuous probability in $[0,1]$ that quantifies the likelihood that flight $i$ will absorb as much delay as is typically feasible at its airport and carrier. Higher values indicate stronger predicted absorption capability, while lower values reflect a greater tendency toward delay propagation or amplification. This learned probability is subsequently incorporated into Stage~II to enhance departure delay prediction.

\subsection{Terminology Clarification: \textit{DelayAbsorbed} vs.\ \textit{AbsorbScore}}

To avoid ambiguity, we distinguish the two key quantities used throughout the framework:

\begin{itemize}
    \item \textbf{DelayAbsorbed} is the binary ground-truth label indicating whether a flight’s realized recovery $\Delta_i$ meets or exceeds the historical buffer $\overline{B}_{a,c}$ of its airport--carrier pair. It provides the supervised learning target for Stage~I.

    \item \textbf{AbsorbScore} is the predicted probability generated by the Stage~I CatBoost classifier. It estimates the flight’s likelihood of achieving successful delay absorption under its observed operational and meteorological conditions.
\end{itemize}

In summary, \textit{DelayAbsorbed} represents a context-normalized measure of actual performance, whereas \textit{AbsorbScore} provides a model-based assessment of expected absorption capability that informs the downstream Stage~II model.

\subsection{Stage II: Predicting Departure Delay}
In the second stage, the AbsorbScore is used as an additional explanatory variable to enhance the prediction of departure delays. We train an XGBoost classifier to estimate the probability of a flight departing with a delay exceeding 15 minutes. The feature vector incorporates flight schedule attributes, operational variables, and meteorological indicators, concatenated with the learned AbsorbScore:
\begin{equation*}
\resizebox{0.9\linewidth}{!}{$
P(\mathrm{DepDelayed}=1 \mid \mathbf{x}, \mathrm{AbsorbScore})
= f_B([\mathbf{x}; \mathrm{AbsorbScore}]; \Theta_B)
$}
\end{equation*}
where $f_B(\cdot)$ denotes the XGBoost model parameterized by $\Theta_B$. To address class imbalance in the delayed flight category, we apply a weighted loss function through the parameter $\texttt{scale\_pos\_weight}$, ensuring that minority-class instances receive proportionally greater emphasis during optimization. 

\begin{figure}[ht]
\centering
\includegraphics[width=0.95\linewidth]{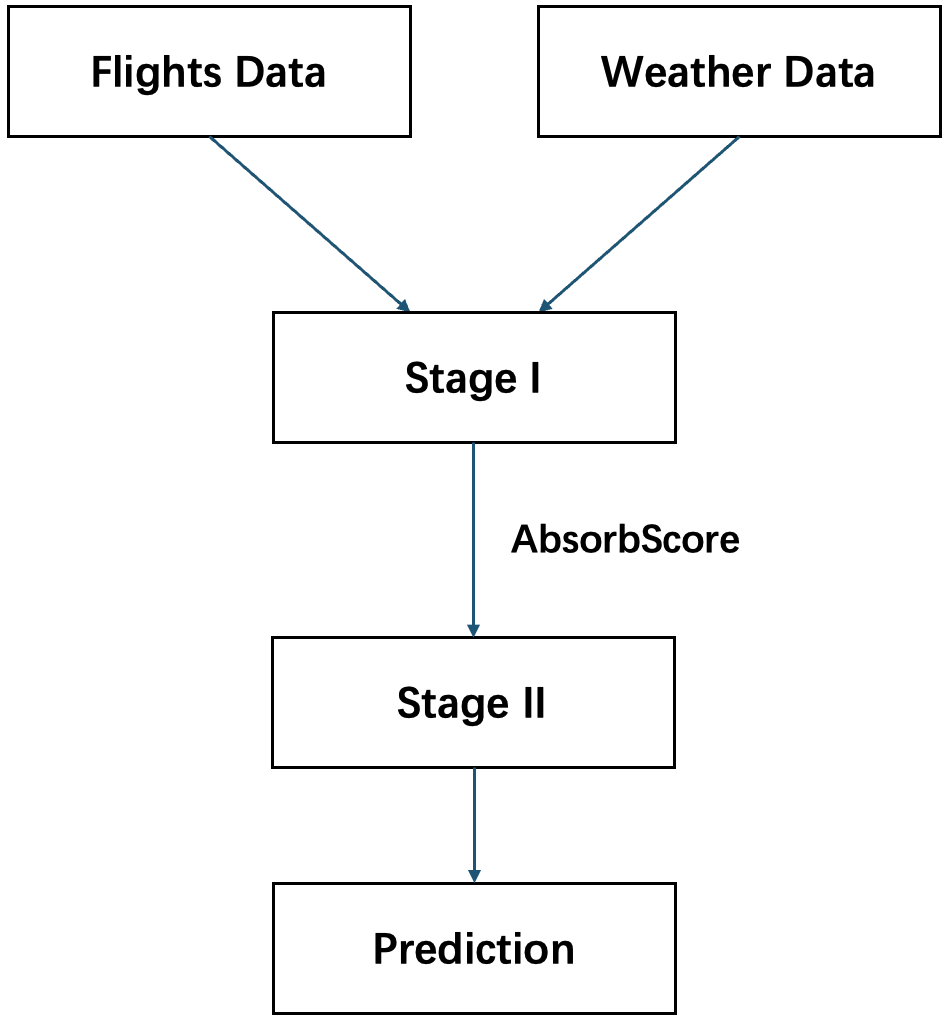}
\caption{Two-stage prediction pipeline: CatBoost predicts DelayAbsorbed as AbsorbScore, then integrate it in XGBoost to predict departure delay.}
\label{fig:model}
\end{figure}

\subsection{Training and Evaluation}

Both stages employ a 5-fold stratified cross-validation scheme and hyperparameter optimization using Optuna to ensure robust generalization and reproducibility. The dataset is partitioned into 80\% for training and 20\% for testing, with all model selection and parameter tuning performed exclusively on the training portion. 

Evaluation metrics include:
\begin{itemize}
    \item \textbf{Area Under the Receiver Operating Characteristic Curve (AUC)} – measures the model’s ability to discriminate between delayed and on-time flights.
    \item \textbf{Average Precision (AP)} – computed as the area under the precision–recall curve, reflecting the model’s performance on the minority (delayed) class.
    \item \textbf{Precision, Recall, and F1-score} – to evaluate classification accuracy under varying thresholds.
    \item \textbf{Confusion Matrix Analysis} – to provide detailed insight into misclassification patterns across both classes.
\end{itemize}

To further enhance the model’s sensitivity to delayed flights, the classification threshold is tuned based on the optimal $F_{\beta}$-score with $\beta > 1$, thereby prioritizing recall for the delayed class. This approach mitigates the imbalance between delayed and on-time flights by emphasizing the correct identification of minority-class instances. Both models are trained using cross-validation, and final predictions are generated on the independent test set. The resulting two-stage architecture improves both predictive performance and interpretability by explicitly linking flight-level delay propagation dynamics with the ultimate departure delay outcome.

\section{Results and Discussion}

\subsection{Model Performance Overview}

To evaluate the proposed modeling framework, four baseline models---Logistic Regression, Decision Tree, Random Forest, and XGBoost---were first implemented using standard numerical and categorical flight features. These models were further compared with the advanced two-stage framework consisting of Stage I (CatBoost for delay absorption capability estimation) and Stage II (XGBoost for departure delay prediction). The classification metrics are summarized in Table~\ref{tab:summary}.

\begin{table}[htbp]
\centering
\caption{Model performance comparison}
\label{tab:summary}
\begin{tabular}{lcc}
\toprule
\textbf{Model} & \textbf{Accuracy} & \textbf{ROC--AUC} \\
\midrule
Logistic Regression & 0.803 & 0.794 \\
Decision Tree & 0.859 & 0.826 \\
Random Forest & 0.861 & 0.851 \\
XGBoost & 0.869 & 0.865 \\
Our model & \textbf{0.892} & \textbf{0.898} \\
\bottomrule
\end{tabular}
\end{table}

The baseline models show that ensemble methods outperform linear and tree-based classifiers, with the XGBoost model achieving the highest performance among the baselines (Accuracy = 0.8686, AUC = 0.8653). This confirms the nonlinear and feature-interactive nature of the flight delay mechanism, which simple linear models fail to capture adequately.

\subsection{Performance of Stage~I: Delay Absorption Capability}
Stage~I was trained to estimate each airport's and aircraft's ability to absorb inbound delays under varying operational and meteorological conditions. When evaluated at the default threshold of 0.5, the CatBoost classifier achieved an accuracy of 0.851, indicating strong discriminative ability between absorbed and propagated delay instances. At the optimized threshold of 0.35 (chosen to maximize the F1-score), Stage~I obtained an F1-score of 0.8866 for delayed cases, better than previous 0.8770 and an overall accuracy of 0.856. The precision--recall trade-off demonstrates that lowering the threshold increases sensitivity to delay absorption, which is desirable when prioritizing operational resilience estimation.

\subsection{Performance of Stage~II: Departure Delay Prediction}
Table~\ref{tab:modelB_metrics} summarizes the classification metrics of Stage~II under two probability thresholds: the conventional 0.5 and the F$_1$-optimized threshold of 0.625. 
At the default threshold, Stage~II achieves an overall accuracy of 0.8975 and a weighted F$_1$-score of 0.882, 
indicating well-balanced performance across both classes. 
When the threshold is tuned to maximize the F$_1$-score, accuracy slightly improves to 0.892, 
suggesting that the classifier is robust and not overly sensitive to threshold selection.

\begin{table}[ht]
\centering
\caption{Performance summary of Stage II under different thresholds.}
\begin{tabular}{lcc}
\hline
\textbf{Metric} & \textbf{Threshold=0.5} & \textbf{Threshold=0.625 (F1-optimal)} \\
\hline
Accuracy & 0.8752 & 0.8917 \\
AUC & 0.8977 & 0.891 \\
AP & 0.828 & 0.829 \\
\hline
Precision (class 0) & 0.9211 & 0.9076 \\
Precision (class 1) & 0.7224 & 0.8218 \\
Recall (class 0) & 0.9169 & 0.9572 \\
Recall (class 1) & 0.7339 & 0.6697 \\
F$_1$-score (class 0) & 0.9190 & 0.9317 \\
F$_1$-score (class 1) & 0.7281 & 0.7380 \\
\hline
\end{tabular}
\label{tab:modelB_metrics}
\end{table}

\begin{figure}
\centering
\includegraphics[width=0.95\linewidth]{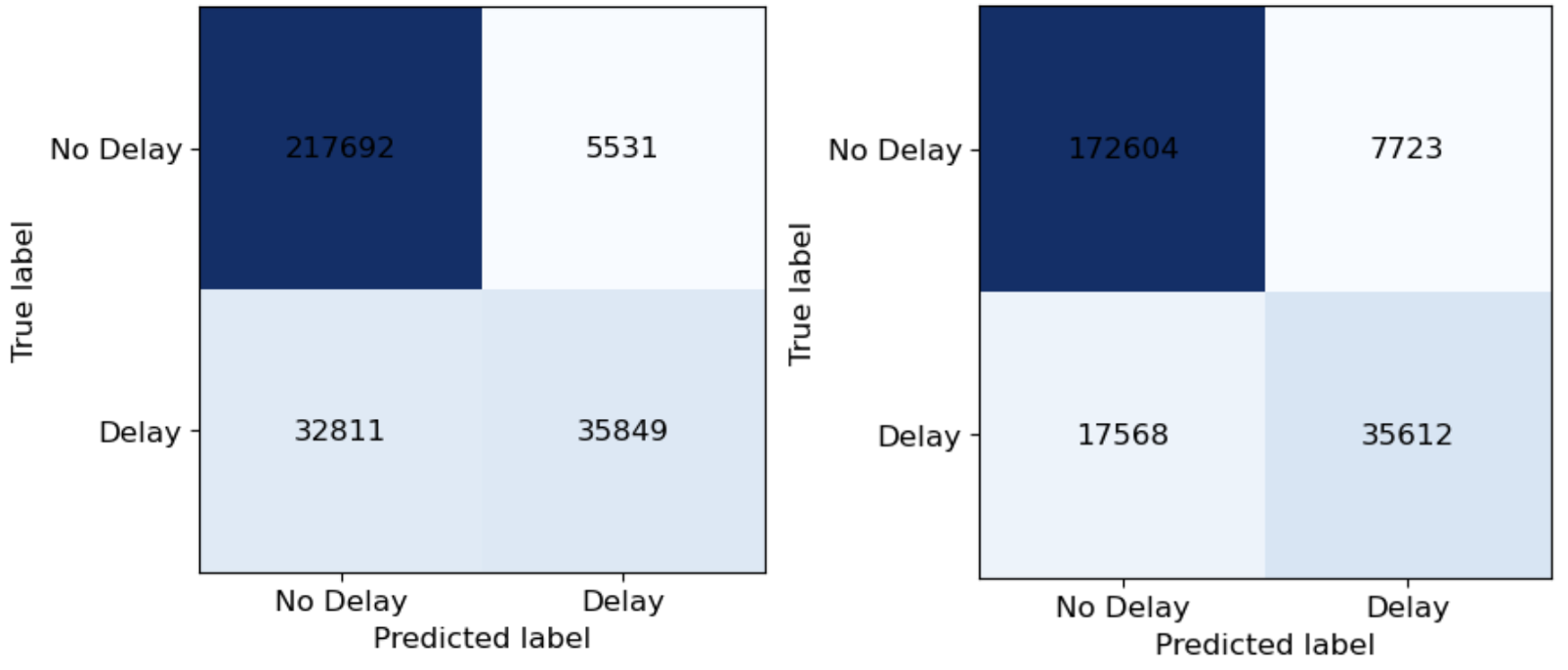}
\caption{Confusion matrices for baseline model XGBoost (left) and our model (right)}
\label{fig:confusion_matrices}
\end{figure}

The confusion matrices in Fig.~\ref{fig:confusion_matrices} reveal that most misclassifications occur in delayed flights (class~1), 
which is expected given the smaller sample proportion and greater variability of delay-causing factors. 
Nevertheless, the model correctly identifies more than 68\% of delayed flights with precision above 82\%, 
demonstrating practical applicability for early warning or operational rescheduling systems.

\subsection{ROC and Precision--Recall Analysis}

\begin{figure}[t]
    \centering
    \includegraphics[width=0.48\textwidth]{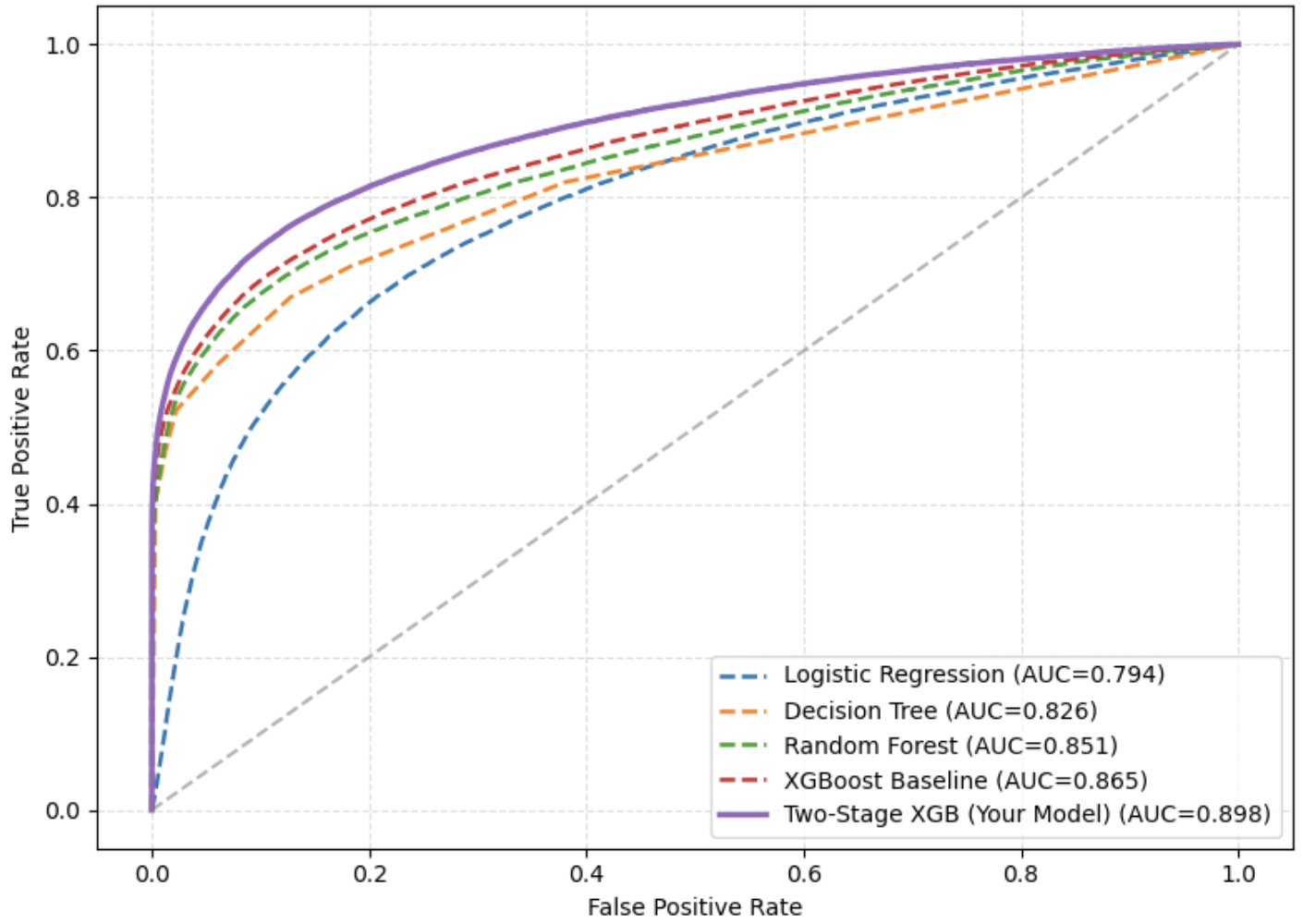}
    \caption{Comparison of ROC curves of different models}
    \label{fig:roc}
\end{figure}

Figure~\ref{fig:roc} presents the ROC curves for our model. The ROC curve exhibits a steep rise near the origin, confirming strong separability between positive and negative samples. An AUC of 0.898 reflects that the model ranks delayed flights above on-time flights nearly 90\% of the time.

\subsection{Feature Importance Analysis}
The feature importance analysis reveals a clear distinction between the two modeling stages. In Stage~I, delay absorption is primarily effected by operational factors such as \textit{PreviousDelay} and \textit{TurnoverTime}, followed by airline and airport-specific effects. These variables collectively capture the physical and organizational mechanisms that determine whether an upstream delay can be recovered during turnaround.

In Stage~II, \textit{AbsorbScore} acts the most influential
predictor, exceeding all other features by an order of magnitude. This demonstrates that learning delay-absorption behavior provides substantial predictive value beyond traditional features such as previous delay, departure hour, and weather. The results validate the two-stage modeling framework: Stage~I effectively quantifies airport- and flight-level resilience, and Stage~II leverages this learned resilience to significantly enhance departure delay prediction accuracy.

\begin{figure}
\centering
\includegraphics[width=0.95\linewidth]{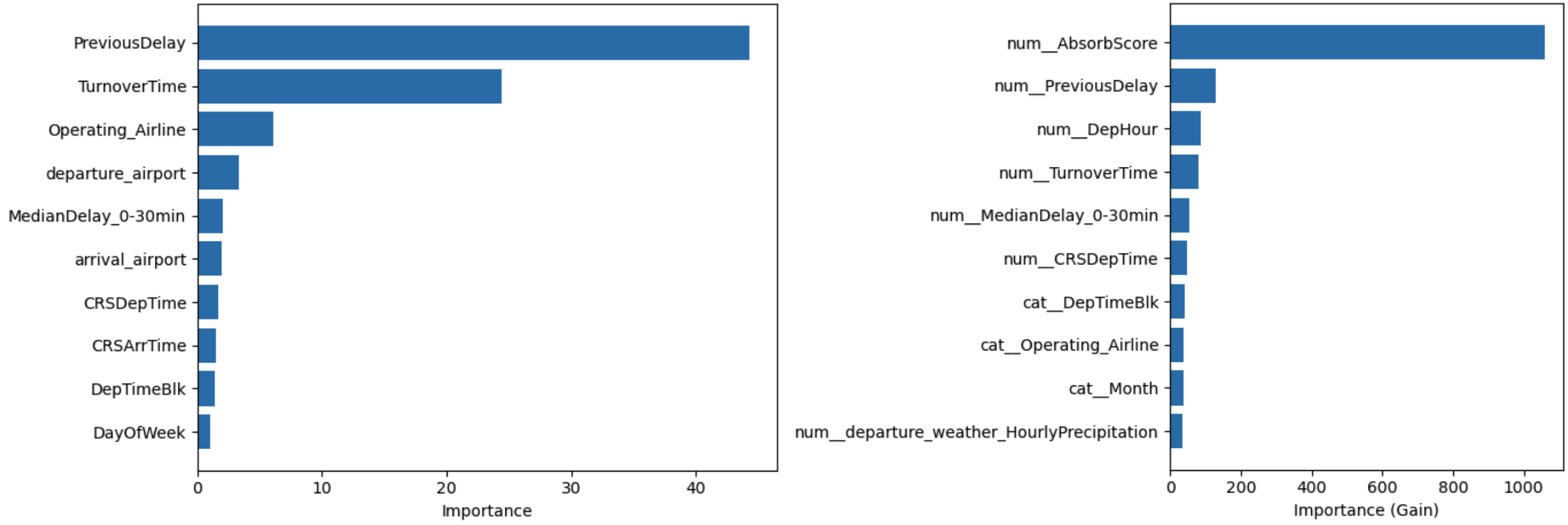}
\caption{The most 10 important features of the two stages of the model}
\label{fig:feature-importance}
\end{figure}

\subsection{Interpretation and Implications}

The AbsorbScore introduced from Stage~I significantly enhances Stage~II’s predictive power. 
By quantifying how efficiently prior delays are absorbed at the aircraft or airport level, 
the model captures temporal and network-level delay propagation that single-stage classifiers often ignore. 
The high precision and AUC confirm that the AbsorbScore contributes interpretable, complementary information to standard operational features.

From an operational standpoint, the results suggest that integrating delay absorption behavior with real-time scheduling and weather data can substantially improve predictive accuracy for short-term delay management. 
Such models can assist dispatchers and controllers in identifying high-risk flights in advance, 
allowing proactive resource allocation, gate adjustments, or crew scheduling to mitigate cascading delays across the network.

\section{Conclusion}
This study introduces a two-stage machine learning framework that incorporates flight-level delay absorption dynamics into departure delay prediction. By explicitly modeling the probability that an inbound delay is absorbed during ground operations, the first-stage CatBoost classifier captures heterogeneous recovery behaviors across airports, airlines, and weather conditions and generates an interpretable AbsorbScore. Incorporating this score into the second-stage XGBoost predictor significantly improves ROC–AUC, precision–recall performance, and delayed-flight identification compared with strong ensemble baselines.

Empirical analyses using U.S. domestic flight and NOAA weather data reveal substantial variations in airports’ delay-absorption capabilities and their influence on downstream delay propagation. Embedding delay-absorption learning into the predictive pipeline enhances both accuracy and operational interpretability, offering actionable value for proactive gate assignment, crew scheduling, and delay-mitigation strategies. Future work may extend this framework to multi-leg rotation modeling, real-time ADS-B or surface-movement integration, and probabilistic assessment of mitigation strategies under adverse weather and congestion.

\bibliographystyle{IEEEtran}
\bibliography{references}

\appendix

\section*{Appendix A: Flight Features From Raw Data}
\begin{itemize}
  \item Year
  \item Quarter
  \item Month
  \item DayofMonth
  \item DayOfWeek
  \item FlightDate
  \item Marketing\_Airline\_Network
  \item Operated\_or\_Branded\_Code\_Share\_Partners
  \item DOT\_ID\_Marketing\_Airline
  \item IATA\_Code\_Marketing\_Airline
  \item Flight\_Number\_Marketing\_Airline
  \item Originally\_Scheduled\_Code\_Share\_Airline
  \item DOT\_ID\_Originally\_Scheduled\_Code\_Share\_Airline
  \item IATA\_Code\_Originally\_Scheduled\_Code\_Share\_Airline
  \item Flight\_Num\_Originally\_Scheduled\_Code\_Share\_Airline
  \item Operating\_Airline
  \item DOT\_ID\_Operating\_Airline
  \item IATA\_Code\_Operating\_Airline
  \item Tail\_Number
  \item Flight\_Number\_Operating\_Airline
  \item OriginAirportID
  \item OriginAirportSeqID
  \item OriginCityMarketID
  \item Origin
  \item OriginCityName
  \item OriginState
  \item OriginStateFips
  \item OriginStateName
  \item OriginWac
  \item DestAirportID
  \item DestAirportSeqID
  \item DestCityMarketID
  \item Dest
  \item DestCityName
  \item DestState
  \item DestStateFips
  \item DestStateName
  \item DestWac
  \item CRSDepTime
  \item DepTime
  \item DepDelay
  \item DepDelayMinutes
  \item DepDel15
  \item DepartureDelayGroups
  \item DepTimeBlk
  \item TaxiOut
  \item WheelsOff
  \item WheelsOn
  \item TaxiIn
  \item CRSArrTime
  \item ArrTime
  \item ArrDelay
  \item ArrDelayMinutes
  \item ArrDel15
  \item ArrivalDelayGroups
  \item ArrTimeBlk
  \item Cancelled
  \item CancellationCode
  \item Diverted
  \item CRSElapsedTime
  \item ActualElapsedTime
  \item AirTime
  \item Flights
  \item Distance
  \item DistanceGroup
  \item CarrierDelay
  \item WeatherDelay
  \item NASDelay
  \item SecurityDelay
  \item LateAircraftDelay
  \item FirstDepTime
  \item TotalAddGTime
  \item LongestAddGTime
  \item DivAirportLandings
  \item DivReachedDest
  \item DivActualElapsedTime
  \item DivArrDelay
  \item DivDistance
  \item Div1Airport
  \item Div1AirportID
  \item Div1AirportSeqID
  \item Div1WheelsOn
  \item Div1TotalGTime
  \item Div1LongestGTime
  \item Div1WheelsOff
  \item Div1TailNum
  \item Div2Airport
  \item Div2AirportID
  \item Div2AirportSeqID
  \item Div2WheelsOn
  \item Div2TotalGTime
  \item Div2LongestGTime
  \item Div2WheelsOff
  \item Div2TailNum
  \item Div3Airport
  \item Div3AirportID
  \item Div3AirportSeqID
  \item Div3WheelsOn
  \item Div3TotalGTime
  \item Div3LongestGTime
  \item Div3WheelsOff
  \item Div3TailNum
  \item Div4Airport
  \item Div4AirportID
  \item Div4AirportSeqID
  \item Div4WheelsOn
  \item Div4TotalGTime
  \item Div4LongestGTime
  \item Div4WheelsOff
  \item Div4TailNum
  \item Div5Airport
  \item Div5AirportID
  \item Div5AirportSeqID
  \item Div5WheelsOn
  \item Div5TotalGTime
  \item Div5LongestGTime
  \item Div5WheelsOff
  \item Div5TailNum
  \item Duplicate
\end{itemize}

\newpage
\section*{Appendix B: Weather Features From Raw Data}
\begin{itemize}
  \item STATION
  \item DATE
  \item LATITUDE
  \item LONGITUDE
  \item ELEVATION
  \item NAME
  \item REPORT\_TYPE
  \item SOURCE
  \item HourlyAltimeterSetting
  \item HourlyDewPointTemperature
  \item HourlyDryBulbTemperature
  \item HourlyPrecipitation
  \item HourlyPresentWeatherType
  \item HourlyPressureChange
  \item HourlyPressureTendency
  \item HourlyRelativeHumidity
  \item HourlySkyConditions
  \item HourlySeaLevelPressure
  \item HourlyStationPressure
  \item HourlyVisibility
  \item HourlyWetBulbTemperature
  \item HourlyWindDirection
  \item HourlyWindGustSpeed
  \item HourlyWindSpeed
  \item Sunrise
  \item Sunset
  \item DailyAverageDewPointTemperature
  \item DailyAverageDryBulbTemperature
  \item DailyAverageRelativeHumidity
  \item DailyAverageSeaLevelPressure
  \item DailyAverageStationPressure
  \item DailyAverageWetBulbTemperature
  \item DailyAverageWindSpeed
  \item DailyCoolingDegreeDays
  \item DailyDepartureFromNormalAverageTemperature
  \item DailyHeatingDegreeDays
  \item DailyMaximumDryBulbTemperature
  \item DailyMinimumDryBulbTemperature
  \item DailyPeakWindDirection
  \item DailyPeakWindSpeed
  \item DailyPrecipitation
  \item DailySnowDepth
  \item DailySnowfall
  \item DailySustainedWindDirection
  \item DailySustainedWindSpeed
  \item DailyWeather
  \item MonthlyAverageRH
  \item MonthlyDaysWithGT001Precip
  \item MonthlyDaysWithGT010Precip
  \item MonthlyDaysWithGT32Temp
  \item MonthlyDaysWithGT90Temp
  \item MonthlyDaysWithLT0Temp
  \item MonthlyDaysWithLT32Temp
  \item MonthlyDepartureFromNormalAverageTemperature
  \item MonthlyDepartureFromNormalCoolingDegreeDays
  \item MonthlyDepartureFromNormalHeatingDegreeDays
  \item MonthlyDepartureFromNormalMaximumTemperature
  \item MonthlyDepartureFromNormalMinimumTemperature
  \item MonthlyDepartureFromNormalPrecipitation
  \item MonthlyDewpointTemperature
  \item MonthlyGreatestPrecip
  \item MonthlyGreatestPrecipDate
  \item MonthlyGreatestSnowDepth
  \item MonthlyGreatestSnowDepthDate
  \item MonthlyGreatestSnowfall
  \item MonthlyGreatestSnowfallDate
  \item MonthlyMaxSeaLevelPressureValue
  \item MonthlyMaxSeaLevelPressureValueDate
  \item MonthlyMaxSeaLevelPressureValueTime
  \item MonthlyMaximumTemperature
  \item MonthlyMeanTemperature
  \item MonthlyMinSeaLevelPressureValue
  \item MonthlyMinSeaLevelPressureValueDate
  \item MonthlyMinSeaLevelPressureValueTime
  \item MonthlyMinimumTemperature
  \item MonthlySeaLevelPressure
  \item MonthlyStationPressure
  \item MonthlyTotalLiquidPrecipitation
  \item MonthlyTotalSnowfall
  \item MonthlyWetBulb
  \item MonthlyAverageWindSpeed
  \item CoolingDegreeDaysSeasonToDate
  \item MonthlyCoolingDegreeDays
  \item MonthlyNumberDaysWithSnowfall
  \item HeatingDegreeDaysSeasonToDate
  \item MonthlyHeatingDegreeDays
  \item MonthlyNumberDaysWithThunderstorms
  \item MonthlyNumberDaysWithHeavyFog
  \item NormalsCoolingDegreeDay
  \item NormalsHeatingDegreeDay
  \item ShortDurationEndDate005
  \item ShortDurationEndDate010
  \item ShortDurationEndDate015
  \item ShortDurationEndDate020
  \item ShortDurationEndDate030
  \item ShortDurationEndDate045
  \item ShortDurationEndDate060
  \item ShortDurationEndDate080
  \item ShortDurationEndDate100
  \item ShortDurationEndDate120
  \item ShortDurationEndDate150
  \item ShortDurationEndDate180
  \item ShortDurationPrecipitationValue005
  \item ShortDurationPrecipitationValue010
  \item ShortDurationPrecipitationValue015
  \item ShortDurationPrecipitationValue020
  \item ShortDurationPrecipitationValue030
  \item ShortDurationPrecipitationValue045
  \item ShortDurationPrecipitationValue060
  \item ShortDurationPrecipitationValue080
  \item ShortDurationPrecipitationValue100
  \item ShortDurationPrecipitationValue120
  \item ShortDurationPrecipitationValue150
  \item ShortDurationPrecipitationValue180
  \item REM
  \item BackupDirection
  \item BackupDistance
  \item BackupDistanceUnit
  \item BackupElements
  \item BackupElevation
  \item BackupEquipment
  \item BackupLatitude
  \item BackupLongitude
  \item BackupName
  \item WindEquipmentChangeDate
  \item Airport
  \item Missing\_Airports
\end{itemize}

\end{document}